# ASKAP-EMU: Overcoming the challenges of wide deep continuum surveys

**Ray P Norris[1]**

*CSIRO Australia Telescope National Facility*
*PO Box 76, Epping, NSW 1710, Australia*
*E-mail:* `Ray.Norris@csiro.au`

**The Emu Team**

*http://askap.pbworks.com/TeamMembers*

Next-generation continuum surveys will be strongly constrained by dynamic range and confusion. For example, the ASKAP-EMU (Evolutionary Map of the Universe) project will map 75% of the sky at 20cm to a sensitivity of 10 μJy – some 45 times deeper than NVSS, and is likely to be challenged by issues of confusion, cross-identification, and dynamic range. Here we describe the survey, the issues, and the steps that can be taken to overcome them. We also explore ways of using multiwavelength data to penetrate well beyond the classical confusion limit, using multiwavelength data, and an innovative outreach approach to cross-identification.

---

[1]  Speaker

## 1. Introduction

Several new radio-telescopes (ASKAP, LOFAR, Meerkat, and ATA) are under construction around the world in the lead-up to SKA. Each has its particular strength, and that of ASKAP (Australian SKA Pathfinder) [5] is to produce wide-deep radio surveys of the sky at around 20 cm wavelength, in both neutral hydrogen and continuum. A process to rank Science Survey Proposals for ASKAP is now complete, with the two highest-ranked projects being the continuum EMU (Evolutionary Map of the Universe) project, and the HI WALLABY (Widefield ASKAP L-band Legacy All-sky Blind survey) project. Each of these will cover 75% of the sky to a depth at least an order of magnitude beyond existing large surveys. Eight other survey projects will also be supported at a lower level than EMU and WALLABY.

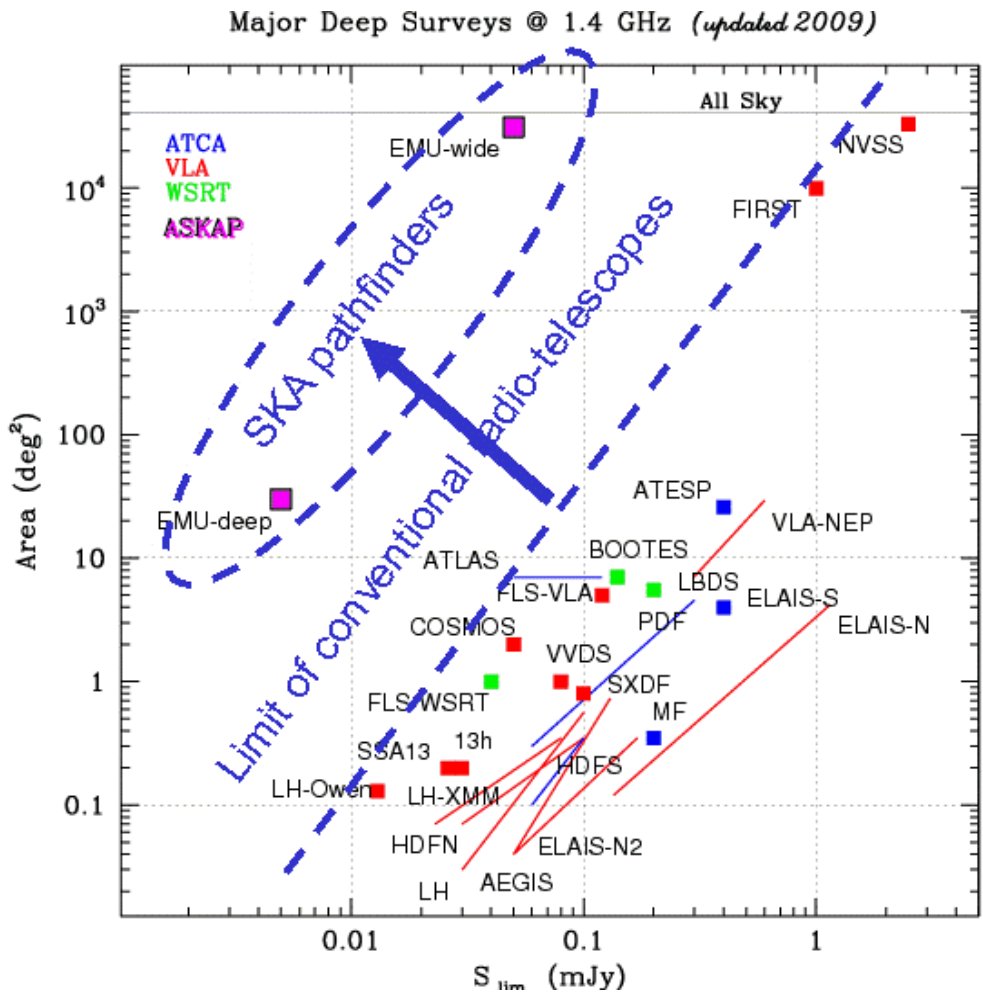

**Fig. 1** Plot showing existing 20-cm surveys, courtesy of Isabella Prandoni. The most sensitive surveys are on the bottom left, and the largest area surveys on the top right. The dotted line shows the limiting envelope of existing surveys, which are largely limited by available telescope time. Next-generation SKA pathfinders such as ASKAP are able to survey the unexplored region of phase space to the top left by using innovative approaches such as focal plane arrays.

The primary goal of EMU is to make a deep (10 μJy rms) radio continuum survey of the entire Southern Sky, extending as far North as +30°. EMU will probe typical star forming galaxies up to z=1, powerful starbursts to even greater redshifts, Active Galactic Nuclei (AGN) to the edge of the Universe, and will undoubtedly uncover new classes of object.

EMU has the potential to exceed the enormous impact that the NRAO VLA Sky Survey (NVSS) [1] had a decade ago, having 45 times more sensitivity (10μJy rms) and an angular resolution (10 arcsec) five times higher than NVSS, whilst covering a similar area (75% of the sky). EMU will survey the entire visible sky to a limit that has currently been achieved in only about 5 sq. deg. distributed among numerous tiny parts of the sky (e.g. the *Hubble* and *Chandra* deep fields), and is expected to yield a catalogue of about 70 million radio sources.

This large area is particularly significant because of the growth in the number of cutting-edge southern hemisphere telescopes and associated planned major surveys spanning all wavelengths. Such a combination of sensitivity and area extends EMU beyond the traditional domain of radio astronomy into the regime previously dominated by UV, optical, and IR surveys. A further difference from NVSS is that the EMU project will include cross-identification with major surveys at other wavelengths (e.g. optical and infrared), many of which are represented in the EMU team.

All radio data from the survey will be placed in the public domain as soon as the data quality has been checked. Cross-identifications of radio sources with other wavelengths will be placed in the public-domain as catalogs accessible using Virtual Observatory (VO) tools.

A separate project, EMU-DEEP, aims to image a single ASKAP pointing with a field of view of 30 sq. deg., to an rms well below the classical confusion limit, using multi-wavelength data to distinguish between confusing sources. EMU-DEEP is not currently supported as an ASKAP Science Survey project, but will be submitted once the technical challenges facing EMU have been successfully addressed.

From 2009 until the survey starts in early 2013, we are conducting a design study to address the challenges that we expect to encounter in the survey. As well as conducting simulations, the design study will use the data currently being obtained in the ATLAS (Australia Telescope Large Area Survey) project, which is imaging 7 sq. deg to an rms of 10μJy [7][6]. Here we describe the issues to be addressed in the design study.

## 2. Science Goals

The key science goals for EMU are:
- To trace the evolution of star-forming galaxies from z=2 to the present day, using a wavelength unbiased by dust or molecular emission.
- To trace the evolution of massive black holes throughout the history of the Universe, and understand their relationship to star-formation.
- To use the distribution of radio sources to explore the large-scale structure and cosmological parameters of the Universe.
- To explore an uncharted region of observational parameter space, almost certainly finding new classes of object.

- To determine how radio sources populate dark matter halos, providing crucial insights into the underlying astrophysics of clusters and halos.
- To create the most sensitive wide-field atlas of Galactic continuum emission yet made in the Southern Hemisphere, addressing areas such as star formation, supernovae, and Galactic structure.

## 3. Technical Challenges to be addressed in the Design Study

### 3.1 Source Density and Confusion

At EMU's 10μJy rms sensitivity, confusion is not expected to be a dominant issue, and this is supported by the observations of HDFS [4], the Phoenix Deep Survey [3], and ATLAS [7] which all reach an rms sensitivity of around 10-20 μJy with a 6-km baseline. We plan to simulate observations of such fields using ASKAP configurations so that we may optimise the weighting function, image processing algorithms, and observing strategy.

### 3.2 Dynamic Range

In each 30-square degree field, we expect ~2 sources > 1 Jy. We hope to reach an rms noise of 10μJy, and so in principle we require a dynamic range of ~$10^5$. However, the 1 Jy sources will not be present in all the sub-beams within the 30-square degrees, and so even if dynamic range limits some sub-beams, we still expect to reach 10μJy rms in others. We will simulate the ASKAP beams to assess the impact of these effects (and others, such as strong diffuse emission from the Galactic Plane) and to determine the optimum weighting of the Focal Plane Array (FPA) feeds, the required uniformity across the beam, the optimum observing strategy, and choice of observing parameters such as weighting.

### 3.3 Image Processing

There is a trade-off between producing naturally weighted (to optimise sensitivity) and uniformly weighted (to optimise resolution) images and the associated source catalogues through the ASKAP pipeline. To extract maximum information from ASKAP, both naturally weighted and uniformly weighted data may be required, implying two processing pipelines for continuum data. Extended structures may require an even more heavily tapered weighting. We plan to simulate data processing using ATLAS data and the ASKAP uv coverage to explore potential solutions, including the use of Briggs robust weighting that may provide an optimal compromise between the two, and also explore the possibility of off-line processing to generate more heavily tapered images than those produced in the real-time pipeline, and to determine the optimal on-line weighting to facilitate this with minimum loss of information.

### 3.4 Source Identification and Measurement

Source extraction algorithms are already under development by the ASKAP software group, but it is extremely difficult to build efficient source extraction algorithms, robust to varying signal-to-noise across the image, and sensitive to source scales ranging from unresolved to extended low-surface-brightness structures. The EMU team includes a considerable body of expertise in this area (e.g. Condon, Hopkins, Huynh, Ivison, Prandoni, and Rudnick) but also a recognition that the radio-astronomical community has not yet achieved an optimum strategy, particularly for extended sources. A major component of the design study will build on our previous experience, and similar developments taking place within LOFAR and other projects, to conduct extensive simulations and develop a robust source detection strategy, which might initially subtract off the obvious point sources and then search for sources at a variety of resolutions using wavelet analysis, genetic algorithms, etc.

**3.5 Cross-Identification**

A key component of the EMU design study is to develop an automated cross-identification pipeline, significantly more sophisticated than a simple nearest-neighbour algorithm, to cross-identify the 70 million radio sources detected by EMU with optical and infrared sources found by other surveys. This design study will leverage the substantial expertise within the EMU team, and build on the experience developed through the international and multiwavelength surveys in which EMU team members are named investigators, and some of whom are developing similar pipelines (e.g. for Herschel).

We also plan Monte-Carlo simulations to derive misidentification rates. During the early part of the Design Study, we will use ATLAS data to investigate possible algorithms, and estimate the computing load. In the later part of the Design Study, we will set up collaborations, ensure access to appropriate surveys, construct software for the pipeline processor, and run simulations to measure the misidentification rates.

**4. An Outreach Approach to Cross-Identification**

Involvement of the public in large data analysis projects was pioneered by SETI@home[10] and subsequently adopted by a large number of other projects. Such outreach programs have evolved from the relatively simple screensaver of SETI@home to GalaxyZoo's (http://galaxyzoo.org/) ambitious goal of harnessing the accumulated pattern-matching ability of thousands of users to classify galaxies from the Sloane Digital Sky Survey.

We plan to explore the further evolution of this trend to the problem of cross-identification of radio sources. The database of 70 million sources generated by EMU will open an opportunity to conduct a public distributed computing project to perform cross-identifications of radio sources against major optical and infrared catalogues and images. During the design study phase, we will explore opportunities to build such a system to engage the public in the process of source classification and cross-identification.

In addition, we plan to explore whether a similar interface can package even more challenging astronomical and astrophysical problems so that they become accessible to intelligent but untutored users. Such an approach has the benefit of providing an intellectual outlet for fertile brains of gifted individuals, perhaps residing in developing countries, who do not have the educational background or opportunities to participate in the intellectual endeavour of astronomy.

# 5. Breaking through the Confusion Barrier: EMU-DEEP

## 5.1 Goals

EMU-DEEP aims to image a single ASKAP pointing to a target depth of 1μJy rms, using multi-spectral information to transcend the classical confusion limit. EMU-DEEP will need 1500 hrs in each of three frequency bands to yield a catalog of about 500,000 sources. The techniques for EMU-DEEP venture into unproven territory, requiring a cautious approach. EMU-DEEP observations will not be proposed until the technical issues facing EMU have all been successfully addressed, and the instrumental characteristics of ASKAP are well-understood.

The key scientific goals for EMU-DEEP are:
- To explore and develop techniques for extracting the maximum information from deep and potentially confused radio images.
- To identify rare or specific populations through techniques involving radio spectral index measurement, prior information based on complementary data, and probabilistic analysis.
- By using such a population of extended radio sources, assess the AGN mechanical feedback and environmental impact on the intergalactic medium, to determine the role it plays in the quenching of galaxy growth in luminous ellipticals at redshift $z=1-2$, and the cosmological evolution in the physical state and distribution of thermal gas in and out of galaxies.

Additional goals, achievable if the challenge of confusion can be overcome, are:
- To trace the evolution of strongly star-forming galaxies from $z=6$ to the present day, and quiescent star forming systems from $z=1$, using a wavelength unbiased by dust or molecular emission.
- To trace the evolution of super-massive black holes throughout the history of the Universe, and understand their relationship to star-formation.
- To explore an uncharted region of observational parameter space, almost certainly finding new classes of object.

Although the target rms of 1μJy is achievable more easily with instruments such as eVLA, such instruments will only be able to image a small area of sky to this depth, and so unusual classes of object will not be detected in such surveys.

## 5.2 Techniques for overcoming the Confusion Limit

### 5.2.1 Radio/IR correlation and Multiwavelength SED modelling

Given a source catalogue at mid- to far-infrared wavelengths (e.g. from Spitzer, Herschel, or WISE), the infrared SED can be used together with the radio-far-infrared (FIR) correlation for star forming galaxies to predict the radio flux densities. Subtracting these source estimates from the radio image will leave a difference image containing those sources (mainly AGNs) that have an SED that differs from star formation in the infrared, together with residuals from the radio-FIR correlation. This approach is likely to be sensitive to variations in the apparent radio-FIR correlation with redshift as a result of k-correction effects [9]. This can be resolved for those infrared sources with known redshifts by applying the appropriate radio-FIR calibration given the redshift. Since at least half the sub-mJy sources are expected to be dominated by star formation, we expect this process to remove about half the sources from the radio image.

### 5.2.2 Radio Spectral Index

Instrumental confusion may also be mitigated by using the fact that a large fraction of radio sources have a synchrotron power law spectrum with spectral index $\alpha = -0.7$ ($S \propto \nu^{\alpha}$). This is the spectral index associated with the star forming galaxy population, expected to account for the majority of the sources in EMU-DEEP, and also with low-luminosity AGNs. Early evidence for average spectral indices to trend toward flatter values at flux densities below 1 mJy may be accounted for by the low luminosity AGN population, which still accounts for the majority of the sources above 0.2 mJy [8]. With observations obtained in two (or more) ASKAP frequency bands, the sources identified in the lowest frequency band can be scaled by an assumed spectral index $\alpha = -0.7$ and subtracted from an image in a higher frequency band. Those sources with an intrinsic spectral index of $\alpha = -0.7$ will be (almost) completely removed, leaving a difference image with a much lower surface density of sources, consisting only of those with flatter or steeper spectral shapes.

This approach will not mitigate instrumental confusion for synchrotron spectrum sources but will allow clear identification of faint populations of the rarer steep (and ultra-steep) and flat spectrum sources. The ultra-steep spectrum subset is of particular interest as such sources are targets for identifying very high redshift radio galaxies[2]. The flat spectrum population will likely be dominated by AGN cores, and will provide a sample for exploring the evolution of low luminosity AGN in the nearby universe and powerful AGN over cosmic history.

### 5.2.3 Low Surface Brightness Sources

Low surface brightness and extended sources may be identified through an approach analogous to that outlined above. By identifying sources in a full resolution image, appropriately smoothed versions can be subtracted from a low resolution image (created either by smoothing the original image, or independently through imaging with different weighting schemes). The residual image will consist primarily of extended and low surface brightness sources. These rare populations form the basis of the investigation into inter-galactic medium and galaxy gaseous halo evolution outlined below. While this approach can be applied to data within a single frequency band, the higher resolution in a high frequency band will facilitate the identification of extended source structures in the lower frequency bands.

## 6. Conclusion

EMU will quantify the evolution of galaxies, identifying the links between star formation and supermassive black holes over the majority of cosmic history. It will have immense legacy value by providing the deepest continuum images possible over three quarters of the entire sky. However, history demonstrates that unexpected discoveries, often of hitherto unknown types of object, are frequently made when a new value of observational phase space is explored for the first time. Perhaps the most significant value of EMU may come from such discoveries.

However, we recognise that a number of technical challenges need to be addressed before we can book our flights to Stockholm.

ASKAP projects are not alone in facing these challenges. Other cutting-edge instruments under construction, such as eVLA, eMERLIN, LOFAR, Meerkat, and ATA, are also confronting similar issues. It is important that all these projects share expertise and development if we are to extend the limits of our existing radio-astronomical techniques and algorithms.